# Experimental observation of optical differentiation and optical Hilbert transformation using a single SOI microdisk chip

Ting Yang[1], Jianji Dong[1*], Li Liu[1], Shasha Liao[1], Sisi Tan[1], Lei Shi[1], Dingshan Gao[1] & Xinliang Zhang[1*]

[1] Wuhan National Laboratory for Optoelectronics, Huazhong University of Science and Technology, Wuhan, China. *Correspondence and requests for materials should be addressed to J.J.D and X.L.Z. (email: jjdong@mail.hust.edu.cn, xlzhang@mail.hust.edu.cn).

# Experimental observation of optical differentiation and optical Hilbert transformation using a single SOI microdisk chip

Optical differentiation and optical Hilbert transformation play important roles in communications, computing, information processing and signal analysis in optical domain which offering huge bandwidth. Meanwhile, silicon-based photonic integrated circuits are preferable in all-optical signal processing due to their intrinsic advantages of low power consumption, compact footprint and ultra-high speed. In this study, we analyze the interrelation between first-order optical differentiation and optical Hilbert transformation and then experimentally demonstrate a feasible integrated scheme which can simultaneously function as first-order optical differentiation and optical Hilbert transformation based on a single microdisk resonator. This finding may motivate the development of integrated optical signal processors.

Optical signal processing has received significant attentions in the past years in an attempt to overcome the bandwidth and speed bottlenecks incurred in conventional electronics[1]. Many optical signal processing systems have been proposed and demonstrated over the years, including all-optical logics[2-5], optical differentiation (OD)[6-11] and integration[12-16], and optical Hilbert transformation (OHT)[17,18]. Particularly, OD has been widely used in optical computing, optical communications, optical metrology, optical digital processing and analog processing, and optical sensing[6-11], while OHT has been used in the applications where signal phase is of critical importance, such as optical phase shifter[17,18], optical single sideband modulation (SSBM)[19,20], vector modulation[21] and frequency measurement[22]. It is very interesting to discover the relationship between OD and OHT since these two functions have some similarities. A theoretical analysis of the interrelation of OD and OHT was discussed in Ref. 18, but no further experimental verification was carried out yet. Since silicon-based photonic integrated circuits are highly desirable in all-optical signal processing due to their intrinsic advantage of low power consumption, compact footprint and ultra-high speed, we attempt to experimentally demonstrate OD and OHT simultaneously based on an on-chip microdisk resonator (MDR). Moreover, the interrelation between OD and OHT is verified in our experiment.

The transfer function of first-order OD can be expressed as[1]

$$H_{OD}(\omega) = j(\omega - \omega_c) \qquad (1)$$

while for an ideal OHT, the transfer function is defined by[17-19]

$$H_{OHT}(\omega) = \begin{cases} \exp(-j\frac{\pi}{2}), & \omega > \omega_c \\ 0, & \omega = \omega_c \\ \exp(+j\frac{\pi}{2}), & \omega < \omega_c \end{cases} \qquad (2)$$

where $j = \sqrt{-1}$, $\omega$ and $\omega_c$ are the optical angular frequency and central angular frequency, respectively. From Eqs. (1) and (2), we can see that the phase responses of the first-order OD and the ideal OHT are the same, with a phase shift of π at the central frequency. However, the amplitude response of the first-order OD is a notch response (linearly varied with the frequency) whereas that of the ideal OHT is an all-pass response, which could not be realized in practice. In fact, when we attempt to obtain an optical filter utilized for OHT, the optical filter normally turns to be a notch filter. Assume that the amplitude response is linearly varied with the frequency near the notch frequency. Then the transfer function is first-order OD at low frequency operation. However, for high frequency operation, the amplitude response can be regarded as a constant. Thus the transfer function satisfies OHT. Therefore, for a given notch filter with a π phase shift at the notch, it is determined by the bandwidth of input signals whether implementing OD or OHT.

To quantitively investigate the interrelation between the first-order OD and OHT, we introduce a deviation analysis during simulation. Assume that the input Gaussian pulse signal has a fixed full width at half maximum (FWHM) of 24.52 ps and a repetition frequency of 6 GHz, as shown in Fig. 1(a). We define a notch filter whose 3-dB bandwidth is tunable, as shown in Fig. 1(b) (blue solid line) and the notch filter has a phase shift from π/2 to -π/2 at the central frequency (shown in Fig. 1(b), red dot line). Make the input Gaussian pulse propagate through the notch filter with its central frequency aligned to the notch. Then the output waveform of the notch filter is calculated, which will be compared with ideal output waveforms of OD and OHT results, respectively.

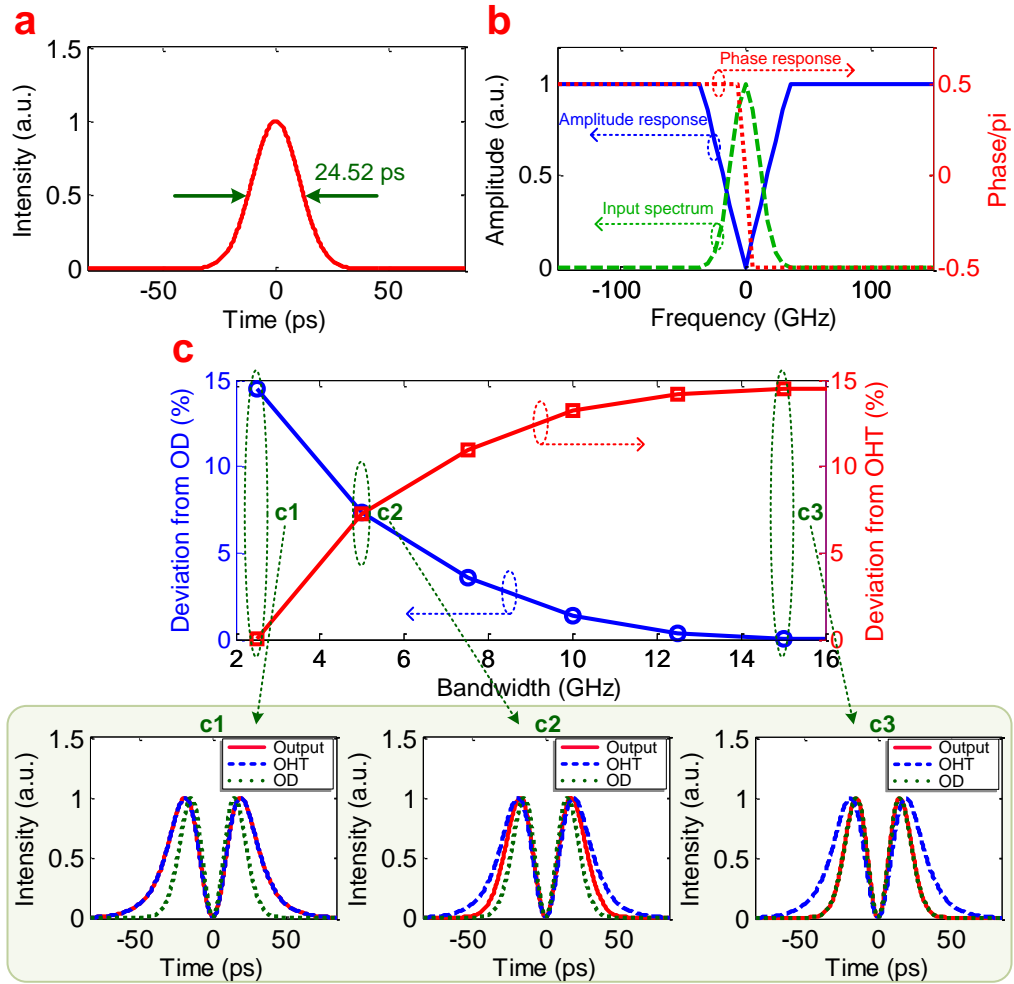

**Figure 1 | Simulation results of the first-order OD and the OHT.** (**a**) input Gaussian pulse with fixed FWHM of 24.52 ps; (**b**) the spectrum of the input Gaussian pulse (green dash line), the amplitude response of the defined filter (blue solid line) with 3-dB bandwidth tunable and the phase response of the defined filter (red dot line); (**c**) the deviations between the output waveform and the OD (blue circle) and the OHT (red square) as a function of the defined filter's 3-dB bandwidth, insets **c1**-**c3** show the output waveforms (red solid line) of the defined filter, ideal OD waveform (green dot line) and ideal OHT waveform (blue dash

line) with different 3-dB bandwidths of the defined filter.

To compare the waveform accuracy quantitively, an average deviation parameter is defined as the mean absolute deviation of the output waveform power from the ideal OD or OHT waveform at certain duration, which is ~166.67 ps in our simulation. Then we vary the 3-dB bandwidth of the notch filter and scan the deviation parameters. Figure 1(c) shows the deviation as a function of the 3-dB bandwidth of notch filter, where blue circle represents the deviation between the output waveform and ideal OD and red solid line represents the deviation between the output waveform and ideal OHT, and the insets c1-c3 show the output waveforms (red solid line) of the notch filter, calculated ideal OD (green dot line) and calculated ideal OHT (blue dash line) with different 3-dB bandwidths of the notch filter. One can see that, for a larger bandwidth of the notch filter, such as 15 GHz, the output waveform accords well with the calculated ideal OD result, otherwise, for a small bandwidth, such as 2.5 GHz, it closely approximates the calculated ideal OHT result. In other words, for a bandwidth-fixed notch filter with a phase shift of $\pi$ at the central frequency, a wideband input signal (small pulsewidth) will result in an OHT result, and a narrowband input signal (large pulsewidth) will result in an OD result on the contrary.

## Results

**Device structure.** As illustrated in Fig. 2(a), a MDR structure fabricated on commercial silicon-on-insulator (SOI) wafer consists of a disk and a bus waveguide. The top silicon thickness of the SOI wafer we used is 340 nm. To form silicon ridge waveguide, the upper silicon layer is etched downward 240 nm, thus the lower silicon layer is 100 nm. Figure 2(b) shows the scanning electron microscope (SEM) image of the fabricated silicon MDR. The radius of the disk is 10 μm, and the waveguide width and thickness of the bus waveguide are 500 nm and 240 nm, respectively. The gap between the disk and the bus waveguide is 150 nm. We use vertical grating coupling method to couple the fiber with the silicon MDR, and the grating coupler we designed has a period of 630 nm with duty cycle of 56%. The coupling loss of the grating coupler is measured to be 5 dB for a single side. Figs. 2(c) and 2(d) show the SEM images of the grating coupler and zoom-in coupling region, respectively.

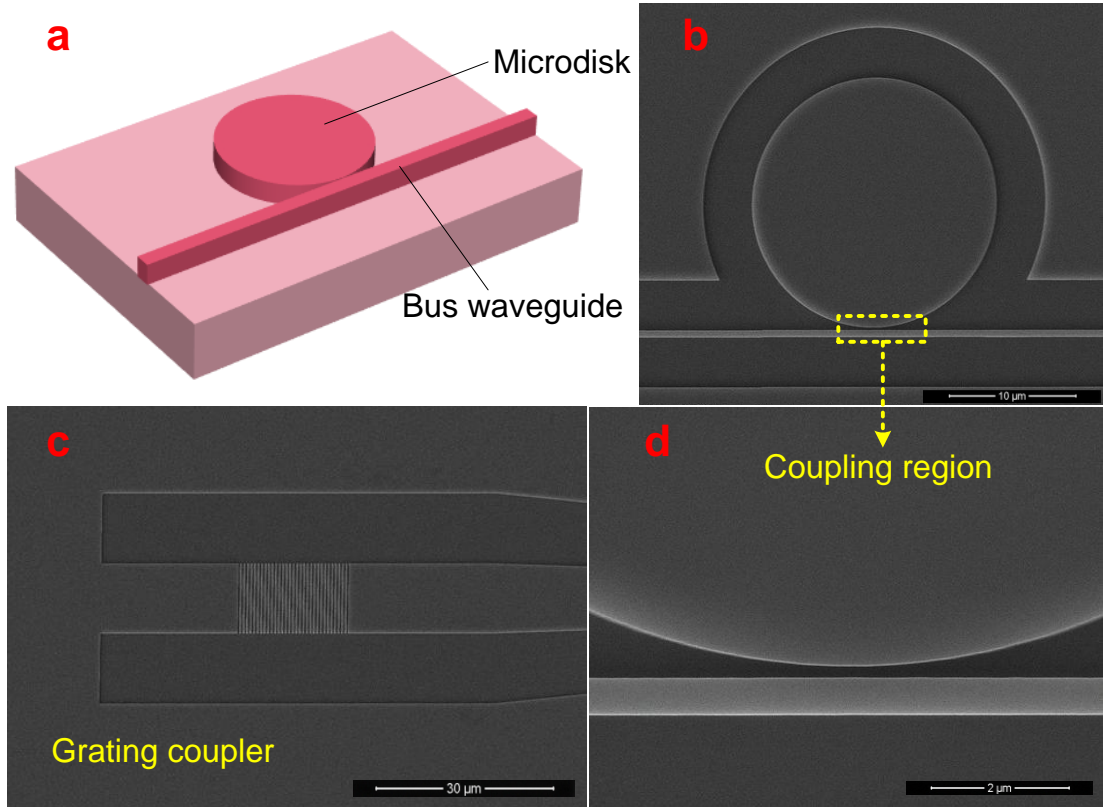

**Figure 2 | MDR design.** (**a**) schematic figure of the MDR composed of a disk and a bus waveguide (vertical grating couplers not shown); (**b**) SEM image of the fabricated MDR, the scale bar represents 10 µm; (**c**) SEM image of the vertical grating coupler, the scale bar represents 30 µm; (**d**) SEM image of the zoom-in coupling region between the disk and the straight waveguide, the scale bar represents 2 µm.

The measured transmission spectrum of the fabricated MDR is illustrated in Fig. 3(a), and Fig. 3(b) shows the zoom-in notch region selected in our experiment, which has a low loss of 12 dB, a relatively high extinction ratio (ER) of 15 dB and a relatively high Q factor of ~13688. The resonant wavelength is around 1554.17 nm. All the spectra are measured by an optical spectrum analyzer (OSA, AQ6370B) with a resolution of 0.02 nm. Figure 3(c) shows the measured phase response of the MDR. One can see a π-phase shift is obtained. The method of phase measurement can be found in Ref. 23.

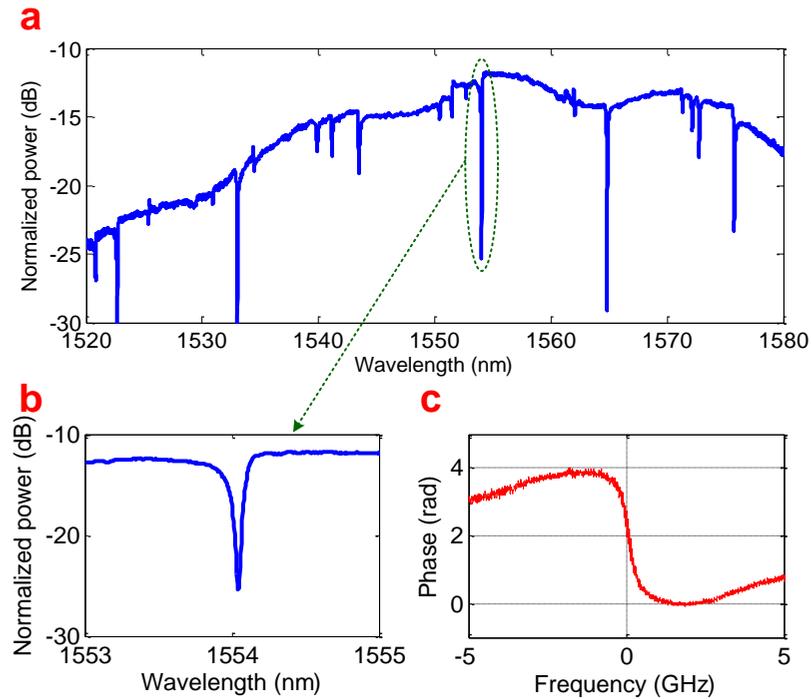

**Figure 3 | Characteristics of the MDR.** (**a**) measured spectrum of the fabricated MDR; (**b**) the zoom-in spectrum of the notch we employed; (**c**) measured phase response of the MDR.

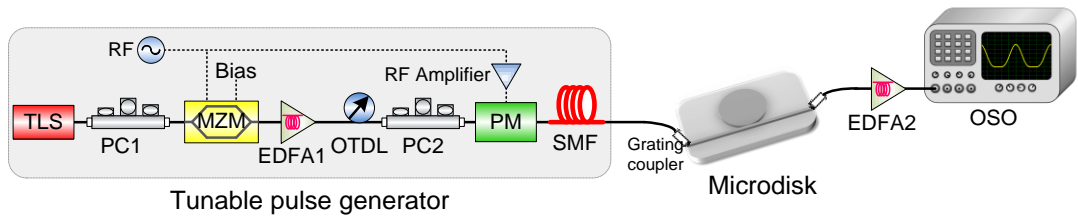

**Figure 4 | Schematic of the experimental setup.** First, a continuous wave beam is emitted by a tunable laser source and is then modulated into Gaussian-like pulses by the tunable pulse generator module. Subsequently, the signal is coupled into the MDR by the designed grating coupler. Finally, the MDR output is amplified and observed. TLS: tunable laser source. PC: polarization controller. RF: radio frequency. MZM: Mach-Zehnder modulator. EDFA: erbium doped fiber amplifier. OTDL: optical tunable delay line. PM: phase modulator. SMF: single mode fiber. OSO: optical sampling oscilloscope.

**Experiment overview.** The experimental setup is shown in Fig. 4. The configuration mainly consists of two parts,

namely, the tunable pulse generator and the MDR. The configuration of tunable pulse generator can be used to generate a Gaussian-like pulse with both repetition frequency and pulsewidth tunable. The output waveform of the MDR is amplified by an erbium doped fiber amplifier (EDFA) to compensate the loss of the MDR and then analyzed through a communication optical sampling oscilloscope (OSO).

**Input Gaussian pulse.** The tunable pulse generator is composed of a tunable laser source (TLS), two polarization controllers (PCs), a Mach-Zehnder modulator (MZM) to produce a direct current (DC) free pulse train, an EDFA to compensate the loss induced by the MZM, an optical tunable delay line (OTDL) used to synchronize the radio frequency (RF) applied on the MZM and phase modulator (PM), a PM to induce large chirp, a segment of single mode fiber (SMF) for chirp compression to generate a short pulse train[24]. By changing the input RF power along with the length of the SMF, pulses with different FWHMs can be obtained. (See Methods for detailed explanation).

**Experimental result.** The measured input and the output temporal waveforms are depicted in Figs. 5(a)-(h), and the calculated waveforms for ideal OD and OHT are also shown for comparison. Figures 5(a), (c), (e) and (g) show the input temporal pulses with FWHMs of 19.57 ps, 24.06 ps, 31.86 ps and 42.53 ps, respectively, fitting by Gaussian pulses. And Figs. 5(b), (d), (f) and (h) are the corresponding measured output waveforms at the MDR output port. One can see that when the pulsewidth of Gaussian pulse is as narrow as 19.57 ps, the output waveform agrees well with the ideal OHT waveform rather than the ideal OD waveform. On the contrary, when the input pulsewidth increases to 42.53 ps, the output waveform accords well with the ideal OD waveform. To analyze the output waveforms, we also calculate the average deviation of the measured output waveform from the calculated theoretical OHT and OD in one certain pulse period. Figure 6 shows the calculated average deviations of input pulse to standard Gaussian pulse (green triangle), output waveform to OHT waveform (red square) and output waveform to OD waveform (blue circle), respectively. With the increase of input pulsewidth, the deviation of output from the calculated theoretical OHT waveform increases, while that of output from the calculated theoretical OD waveform decreases. This trend has a good agreement with the theoretical analysis in Fig. 1(c).

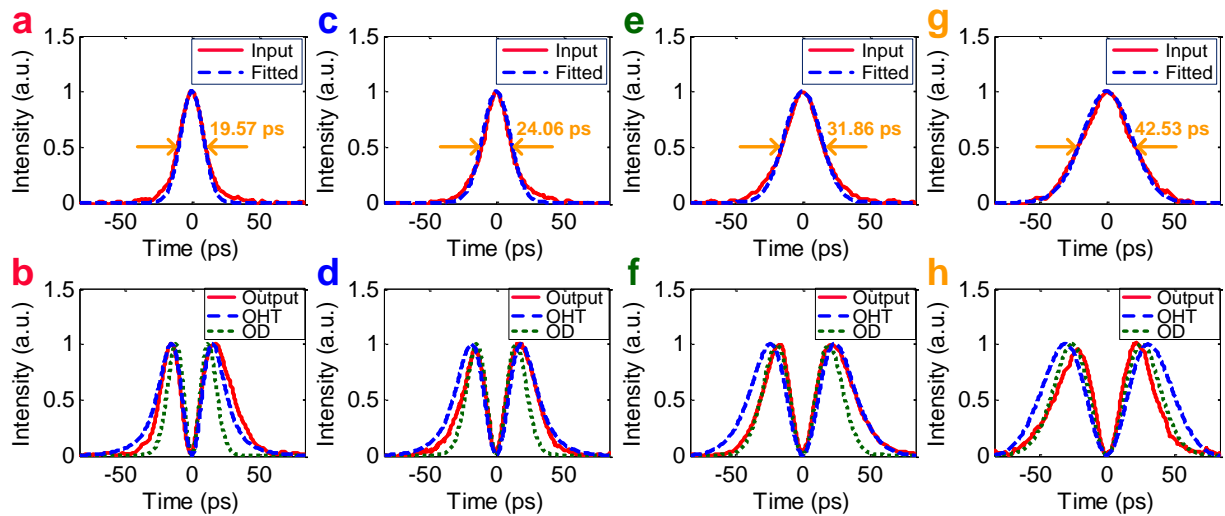

**Figure 5 | Experimental results.** (**a**), (**c**), (**e**) and (**g**) are the inputs; red solid lines: input pulses, blue dash lines: fitted Gaussian pulses. (**b**), (**d**), (**f**) and (**h**) are the outputs; red solid lines: output waveforms, blue dash lines: calculated OHT, green dot lines: calculated OD.

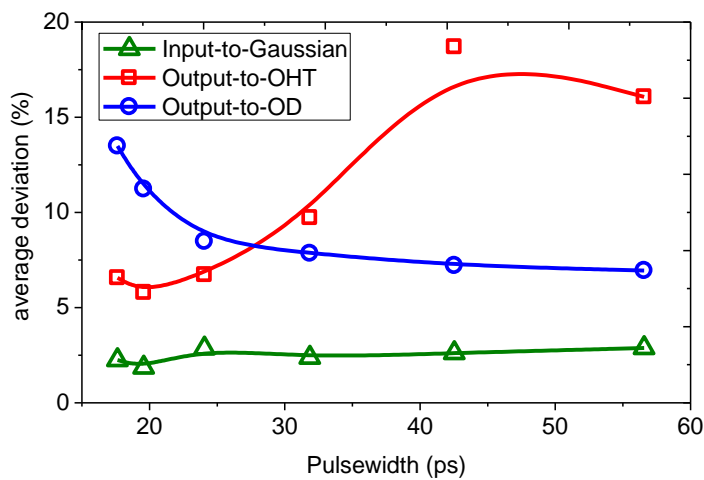

**Figure 6 | Analysis of the temporal waveforms of the MDR.** Green triangle: average deviation between the input pulse and the fitted Gaussian pulse, red square: average deviation between the output waveform and the calculated theoretical OHT waveform, blue circle: average deviation between the output waveform and the calculated theoretical OD waveform.

## Discussion

In this work, we observed both the OD and OHT in a single SOI MDR chip. First, we numerically analyzed the interrelation between OD and OHT, as illustrated in Fig. 1. And we found that, for a properly designed notch filter, it can simultaneously function as OD and OHT under different input conditions. Because of a relatively high Q factor and ~π phase shift at the resonated frequency, a properly designed MDR was chosen to demonstrate the OD and OHT, and the measured spectrum of the fabricated MDR was shown in Fig. 3. Then we performed the experiment. As illustrated in Figs. 5 and 6, we confirmed in experiments that a single SOI MDR can function as an OD or an OHT as the input pulsewidth varied. Because of limited bandwidth of the MDR, the output waveform deviates from the ideal OD result under a short pulse injection. Additionally, due to practically gradual transition of amplitude response at the notch, the output waveform deviates from the ideal OHT result under a wide pulse injection. All the experimental results accord well with the simulated results.

In the part of Gaussian pulse generation, some chirps are induced by the PM firstly and then are compensated by the SMF. The input Gaussian pulse may still carry small chirps due to the incomplete compensation. Thus we need to analyze the impact of chirped pulse during the OD or OHT. From theoretical analysis, we find that whether the input Gaussian pulse is chirped or not, the calculated OD waveform remains the same. For the OHT, the output waveform may have very tiny distortion with asymmetry if the chirp parameter of the input Gaussian pulse is small. In the experiment, once we change the input RF power, we then try to generate a shortest pulse with constantly optimizing the length of SMF. Thus, it is reasonable to deduce that the chirp of the input Gaussian pulse is very small, without severe impacts on the output results.

In summary, we have demonstrated an integrated scheme which can simultaneously function as OD and OHT based on a single SOI MDR. Also we have experimentally verified the interrelation between the OD and OHT. The scheme shows the advantages of compactness, flexibility, versatility and expandability due to the advantage of silicon-based technology.

## Methods

**Input Gaussian pulse.** We used a tunable laser source with a tuning resolution of 0.01 nm, which enables us to precisely align the signal wavelength to the resonance peak of MDR that we employ. The continuous wave (CW) light from the TLS is first modulated by the MZM with sinusoidal RF signal driven. To obtain a DC free pulse from the MZM, the driving parameters of the MZM should be carefully adjusted. Utilizing the nonlinear transmission curve of the MZM, a return-to-zero (RZ) pulse will be obtained when the MZM DC bias is adjusted to below the quadrature point and the amplitude of RF signal is less than $V_\pi$ [25]. In this way, the pulse duty cycle can be less than 50%. In order to induce large frequency chirp to the pulse, an RF amplifier is employed to amplify the RF signal driven on PM. Then a segment of SMF is used to compensate the chirp induced by the PM. It is known that as the SMF length increases, the FWHM is decreased to the minimum, and then is increased subsequently. To get a shortest pulse, we need to optimize the SMF length so that the pulse chirp can be totally compensated by the SMF dispersion. In the experiment, by changing the input RF power along with the length of the SMF, pulses with different FWHMs can be obtained.

**Devices fabrication.** We employ an on-chip MDR to observe first-order OD and OHT. First we design and fabricate the MDR on an SOI wafer. The thickness of the top silicon and the buried oxide layer of the SOI wafer are 340 nm and 2 μm, respectively. Then the device layout is transferred to ZEP520A photoresist by E-beam lithography (Vistec EBPG5000+ES). Then the upper silicon layer is etched downward for 240 nm to form silicon ridge waveguide and input/output grating couplers, through inductively coupled plasma (ICP) etching (Oxford Instruments Plasmalab System 100). We use vertical grating coupling method to couple the fiber and the silicon

MDR, and the grating coupler we designed has a period of 630 nm with duty cycle of 56%. The coupling loss of the grating coupler is measured to be 5 dB for a single side[26].

**References**

1.  Azaña, J. Ultrafast Analog All-optical Signal Processors Based on Fiber-Grating Devices. *Photonics Journal* **2**, 359-386 (2010).

2.  Robinson, A. L. Multiple quantum wells for optical logic. *Science* **225**, 822-824 (1984).

3.  Smith, S., Walker, A., Tooley, F. & Wherrett, B. The demonstration of restoring digital optical logic. *Nature* **325**, 27-31 (1987).

4.  Wei, H., Wang, Z., Tian, X., Käll, M. & Xu, H. Cascaded logic gates in nanophotonic plasmon networks. *Nature Communications* **2**, 387 (2011).

5.  Lei, L., Dong, J., Yu, Y., Tan, S. & Zhang, X. All-Optical Canonical Logic Units-Based Programmable Logic Array (CLUs-PLA) Using Semiconductor Optical Amplifiers. *Journal of Lightwave Technology* **30**, 3532-3539 (2012).

6.  Ngo, N. Q., Yu, S. F., Tjin, S. C. & Kam, C. H. A new theoretical basis of higher-derivative optical differentiators. *Opt. Commun.* **230**, 1-3 (2004).

7.  Kulishov, M. & Azaña, J. Long-period fiber gratings as ultrafast optical differentiators. *Optics letters* **30**, 2700-2702 (2005).

8.  Slavík, R., Park, Y., Kulishov, M., Morandotti, R. & Azaña, J. Ultrafast all-optical differentiators. *Optics Express* **14**, 10699-10707 (2006).

9.  Xu, J., Zhang, X., Dong, J., Liu, D. & Huang, D. High-speed all-optical differentiator based on a semiconductor optical amplifier and an optical filter. *Optics letters* **32**, 1872-1874 (2007).

10. Berger, N. K. et al. Temporal differentiation of optical signals using a phase-shifted fiber Bragg grating. *Opt. Express* **15**, 371-381 (2007).

11. Liu, F. et al. Compact optical temporal differentiator based on silicon microring resonator. *Opt. Express* **16**, 15880-15886 (2008).

12. Quoc Ngo, N. Design of an optical temporal integrator based on a phase-shifted fiber Bragg grating in transmission. *Optics letters* **32**, 3020-3022 (2007).

13. Asghari, M. H. & Azaña, J. Proposal for arbitrary-order temporal integration of ultrafast optical signals using a single uniform-period fiber Bragg grating. *Optics letters* **33**, 1548-1550 (2008).

14. Park, Y. & Azaña, J. Ultrafast photonic intensity integrator. *Optics letters* **34**, 1156-1158 (2009).

15. Ferrera, M. *et al.* On-chip CMOS-compatible all-optical integrator. *Nature Communications* **1**, 29 (2010).

16. Ferrera, M. *et al.* All-optical 1st and 2nd order integration on a chip. *Optics express* **19**, 23153-23161 (2011).

17. Asghari, M. H., & Azaña, J. All-optical Hilbert transformer based on a single phase-shifted fiber Bragging grating: design and analysis. *Optics Letters* **34**, 334-336 (2009).

18. Ngo, N. Q. & Song, Y. On the interrelations between an optical differentiator and an optical Hilbert transformer. *Optics Letters* **36**, 915-917 (2011).

19. Emami, H., Sarkhosh, H., Bui, L. A. & Mitchell, A. Wideband RF photonic in-phase and quadrature-phase generation. *Optics Letters* **33**, 98-100 (2008).

20. Kawanishi, T. & Izutsu, M. Linear single-sideband modulation for high-SNR wavelength conversion. *IEEE Photon. Technol. Lett.* **16**, 1534-1536 (2004).

21. Bui, L. A., Mitchell, A., Ghorbani, K. & Chio, T.-H. Wide-band RF photonic second order vector sum phase-shifter. *IEEE Microw. Wirel. Compon. Lett.* **15**, 309-311 (2013).

22. Nguyen, L.V.T. & Hunter, D.B. A photonic technique for microwave frequency measurement. *IEEE Photon. Technol. Lett.* **18**, 1188-1190 (2006).

23. Tang, Z., Pan, S. and Yao, J. A high resolution optical vector network analyzer based on a wideband and wavelength-tunable optical single-sideband modulator. *Opt. Express* **20**, 6555-6560 (2012).

24. Yang, T., Dong, J., Liao, S., Huang, D. & Zhang, X. Comparison analysis of optical frequency comb generation with nonlinear effects in highly nonlinear fibers. *Optics Express.* **21**, 8508-8520 (2013).

25. Ji, Y., Li, Y., Li, W., Hong, X., Guo, H., Zuo, Y., Zhang, Xu, K., Wu, J. & Lin, J. Generation of 40 GHz phase stable optical short pulses using intensity modulator and two cascaded phase modulators. *Front. Optoelectron. China.* **4**, 292-297 (2011).

26. Dong, J., Zheng, A., Gao, D., Liao, S., Lei, L., Huang, D. and Zhang X. High-order photonic differentiator emplying on-chip cascaded microring resonators. *Optics letters* **38**, (2013).

## Acknowledgement

This work is partially supported by the National Basic Research Program of China (Grant No. 2011CB301704), the Program for New Century Excellent Talents in Ministry of Education of China (Grand No. NCET-11-0168), a Foundation for Author of National Excellent Doctoral Dissertation of China (Grand No. 201139), and the National Natural Science Foundation of China (Grand No. 11174096). The authors would like to thank Dr. Qingzhong Huang and Prof. Jinsong Xia in the Center of Micro-Fabrication and Characterization (CMFC) of Wuhan National Laboratory for Optoelectronics (WNLO) for the assistance in device fabrication, and thank the facility support of the Center for Nanoscale Characterization and Devices, WNLO.

## Author contributions

## Additional information

**Competing financial interests:** The authors declare no competing financial interests.